# Stability of Scheduled Message Communication over Degraded Broadcast Channels

KCV Kalyanarama Sesha Sayee, Utpal Mukherji
Dept. of Electrical Communication Engineering
Indian Institute of Science, Bangalore-560012, India
Email: sayee, utpal@ece.iisc.ernet.in

*Abstract*—We consider scheduled message communication over a discrete memoryless degraded broadcast channel. The framework we consider here models both the random message arrivals and the subsequent reliable communication by suitably combining techniques from queueing theory and information theory. The channel from the transmitter to each of the receivers is quasi-static, flat, and with independent fades across the receivers. Requests for message transmissions are assumed to arrive according to an i.i.d. arrival process. Then, (i) we derive an outer bound to the region of message arrival vectors achievable by the class of stationary scheduling policies, (ii) we show for any message arrival vector that satisfies the outerbound, that there exists a stationary "state-independent" policy that results in a stable system for the corresponding message arrival process, and (iii) under two asymptotic regimes, we show that the stability region of nat arrival rate vectors has information-theoretic capacity region interpretation.

## I. INTRODUCTION

Multi-access random-coded communication with independent decoding, of messages that arrive in a Poisson process to an infinite transmitter population, and that achieves any desired value for the upper bound by determining message signal durations appropriately, has been considered in [1] and [2]. Recently, in [3], a generalization and extension of the model in [1] and [2] was considered and the following assertions were proved: (i) in the limit of large message alphabet size, the stability region has an interference limited information-theoretic capacity interpretation, (ii) state-independent scheduling policies achieve this asymptotic stability region, and (iii) in the asymptotic limit corresponding to immediate access, the stability region for non-idling scheduling policies is shown to be identical irrespective of received signal powers. The work reported in [3] is followed in [4], considering joint decoding of messages, instead of independent decoding. As such, this paper is a sister paper to our discussion of multi-access message communication with independent decoding [3] and joint decoding [4].

In this paper we consider message (packet) communication over a flat bandpass AWGN broadcast channel with $J \geq 2$ receivers. Requests for message transmissions to different receivers are generated according to i.i.d. processes. Requests intended for receiver-$j$, $1 \leq j \leq J$, are chosen from the message alphabet $\mathcal{M}_j$ consisting of $M_j \geq 2$ alternatives. Signals, representing messages, are to be communicated reliably; reliability required by the $j$th receiver is quantified by the tolerable message decoding error probability $p_{ej}$. We assume that the transmitter schedules messages for transmission, i.e., the transmitter can choose some numbers of messages meant for each of the $J$ receivers and then perform superposition encoding [5] on them. Due to the complexity involved in superposition encoding of an arbitrary number of messages, we restrict the transmitter to encode only a finite number $\mathsf{K} \geq 1$ of messages at a time. This restriction gives rise to a set of possible schedules $\mathcal{S}_\mathsf{K}$ defined in the Section II. The channel from the transmitter to each of the receivers is a discrete-time memoryless channel with known statistics that remain stationary over time. The actual communication is accomplished as follows. For a chosen schedule $s \in \mathcal{S}_\mathsf{K}$, the transmitter maps the schedule $s$ to a codeword (signal) of length $N(s)$ and then broadcasts the signal. The length of the code word is carefully chosen so that reliable communication for each receiver, quantified by $\{p_{ej}; 1 \leq j \leq J\}$, is achieved. Decoders, at the respective receivers, perform successive decoding on their received signals and map to an estimate of the messages intended for them.

The contributions in this paper are as follows. We derive an outer bound $\mathcal{R}_{out}$ to the stability region of message arrival rate vectors $\mathbb{E}A = (\mathbb{E}A_1, \mathbb{E}A_2, \ldots, \mathbb{E}A_J)$ achievable by the class of stationary scheduling policies. Next, we propose a class of stationary policies, called "*state-independent*" policies, and then characterize the stability region $\mathcal{R}(\omega)$ of message arrival rate vectors $\mathbb{E}A = (\mathbb{E}A_1, \mathbb{E}A_2, \ldots, \mathbb{E}A_J)$ achievable by any such policy $\omega$. We then go on to establish that for any message arrival rate vector that satisfies the outerbound derived for the stationary policies, there exists a state-independent scheduling policy $\omega$ that results in a stable system for the corresponding message arrival process. Finally, under two asymptotic regimes, we give information-theoretic capacity region interpretation to the stability region of nat arrival rate vectors achievable by a fixed schedule $s \in \mathcal{S}_\mathsf{K}$.

The organization of the paper is as follows. Section II introduces the information-theoretic model of degraded broadcast channel to be analyzed in this paper. We extend a random coding bound derived for a two receiver model [6] to an arbitrary number of receivers. Section III gives a queueing system model for the degraded broadcast message communication system with random message arrivals by characterizing service requirement of messages and the service process of an equivalent server. Section IV presents an outer bound to the stability region of message load vectors achievable by the class of stationary scheduling policies. In section V, we give stability analysis of the queueing model for the class of state-independent scheduling policies. Finally, in Section VI, we give information-theoretic capacity region interpretation to the stability region of message average nat arrival rate vectors.

## II. THE INFORMATION THEORETIC MODEL

The capacity region for general degraded broadcast channels, first conjectured in [7], was established by Bergmans [5]. The converse was established by Bergmans [8] and Gallager [6]. The model for a degraded broadcast channel with $J$ receivers is shown in Fig. 1. Consider a degraded broadcast channel through which $J$ independent sources communicate information to the respective receivers. In Fig. 1 we note that, for $2 \leq j \leq J$, $j$th channel is degraded version of $j-1$th channel.

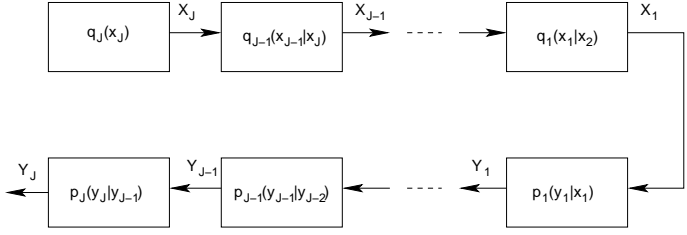

Fig. 1. Model of Degraded Broadcast Channel

*Theorem 2.1 (Bergmans):* The capacity region for the degraded broadcast channel consisting of $J$ component channels (receivers) and represented as the Markov chain $X_J \to X_{J-1} \to \ldots \to X_2 \to X_1 \to Y_1 \to Y_2 \to \ldots \to Y_{J-1} \to Y_J$ is the convex hull of the closure of all $(R_1, R_2, \ldots, R_J) \in \mathbb{R}_+^J$ satisfying $R_J \leq I(X_J; Y_J)$ and for $1 \leq j \leq J-1$, $R_j \leq I(X_j; Y_j | X_{j+1})$ for some joint distribution $q_J(x_J)q_{J-1}(x_{J-1}|x_J)\cdots q_1(x_1|x_2)p(y_1y_2\cdots y_J|x_1)$ ∎

For $1 \leq j \leq J$ and integers $M_j \geq 2$, let $\mathcal{M}_j = \{1, 2, \ldots, M_j\}$ denote the message alphabet for the $j$th source. Let the $j$th source output be modeled by the random variable $m_j$ that takes values in the set $\mathcal{M}_j$. The ensemble of broadcast codes we consider here is the same as Bergmans [5] constructed. For $j \leq k \leq J$, let $\hat{m}_{k,j} \in \mathcal{M}_j$ denote an estimate of the $k$th source computed at the $j$th receiver.

The equivalent baseband channel between the transmitter and each of the $J$ receivers can be sampled at the Nyquist rate equal to its two-sided bandwidth $W$, to obtain a sequence of single-use scalar channels. These channels $ij$, the $i$th scalar channel corresponding to the channel between the transmitter and the $j$th receiver, have independent inputs of variance $|h_j|^2 P$, where $P$ denotes average power of the transmitter and $h_j$ the multiplicative gain. Let $\sigma_j^2$ denote the variance of the additive Gaussian noise. If $|h_1|^2/\sigma_1^2 \geq |h_2|^2/\sigma_2^2 \geq \ldots |h_J|^2/\sigma_J^2$, then the broadcast channel is stochastically degraded. But, in what follows, we analyze a general degraded broadcast channel without any specific model in mind. Thus the results obtained here apply to flat bandpass AWGN broadcast channels. A random coding upper bound for the two receiver degraded broadcast channel was derived in [6]. Here we extend that result to a degraded broadcast channel with arbitrary number of receivers. The objective of the decoder at the $j$th receiver is to compute an estimate $\hat{m}_{j,j}$ of $m_j$. This is achieved by successive decoding, with the $j$th decoder first decoding and then subtracting the signals intended for the users with noisier channels before decoding its own. Let the event $\{\hat{m}_{k,j} \neq m_k\}$ be the event that decoder at the $j$th receiver makes error in decoding the $k$th source. The probability of error for the $j$th decoder then is $p(\{\hat{m}_{j,j} \neq m_j\})$. For $1 \leq j \leq J$ and $j \leq k \leq J$, let $p_{e,k,j}$ denote the expected probability, over the ensemble of broadcast codes, of decoding the $k$th source at the $j$th receiver *incorrectly* conditioned on $k+1, k+2, \ldots, J$th sources being decoded *correctly*. The transition probability of the effective channel between $Y_j$, $1 \leq j \leq J$, and $X_k$, $1 \leq k \leq J$, is given by

$$p'_{Y_j|X_k}(y_j|x_k) = \left(\prod_{l=1}^{k-1} q_{k-l}(x_{k-l}|x_{k-l+1})\right)p(y_1|x_1)$$
$$= \left(\prod_{l=2}^{j} p_l(y_l|y_{l-1})\right)$$

One can then think that $Y_j$ is produced by passing $X_k$ through a DMC with transition probability law $p'_{Y_j|X_k}(y_j|x_k)$. In the following Theorem 2.2, we compute an upper bound on probability of the event $\{\hat{m}_{j,j} \neq m_j\}$.

*Theorem 2.2:* For $1 \leq j \leq J$, the expected error probability over the ensemble of broadcast codes of length $N$ satisfies $p(\{\hat{m}_{j,j} \neq m_j\}) \leq \sum_{k=j}^{J} p_{e,k,j}$, where for $j \leq k \leq J-1$,

$$\begin{aligned}
p_{e,k,j} &\leq \exp(-N E_{X_k,Y_j}(R_k)) \\
E_{X_k,Y_j}(R_k) &= E_{o,X_k,Y_j}(\rho) - \rho R_k \\
E_{o,X_k,Y_j}(\rho) &= -\ln \sum_{x_J,\ldots,x_{k+1}} q_J(x_J) \prod_{l=k+1}^{J-1} q_l(x_l|x_{l+1}) \\
&\quad \sum_{y_j} \left(\sum_{x_k} q_k(x_k|x_{k+1}) p'_{Y_j|X_k}(y_j|x_k)^{\frac{1}{1+\rho}}\right)^{1+\rho} \\
R_k &= \frac{\ln M_k}{N}
\end{aligned} \quad (1)$$

and for $k = J$,

$$\begin{aligned}
p_{e,J,j} &\leq \exp(-N E_{X_J,Y_j}(R_J)) \\
E_{X_J,Y_j}(R_J) &= E_{o,X_J,Y_j}(\rho) - \rho R_J i \\
E_{o,X_J,Y_j}(\rho) &= -\ln \sum_{y_j} \left(\sum_{x_J} q_J(x_J) p'_{Y_j|X_J}(y_j|x_J)^{\frac{1}{1+\rho}}\right)^{1+\rho} \\
R_J &= \frac{\ln M_J}{N}
\end{aligned} \quad (2)$$

∎

In what follows we allow for the possibility of scheduling *multiple* messages intended for a receiver. Let $s = (s_1, s_2, \ldots, s_J) \in \mathbb{Z}_+^J$, a vector of non-negative integers, define a schedule. Then the set $\mathcal{S}_{\mathsf{K}} = \left\{s : 0 \leq \sum_{j=1}^{J} s_j \leq \mathsf{K}\right\}$ defines the set of all schedules that schedule at most $\mathsf{K}$ messages for encoding. To interpret Theorem 2.2 for the schedule $s \in \mathcal{S}_{\mathsf{K}}$, it is convenient to view schedule $s$ as defining new message alphabets for receivers that are product versions of their original message alphabets. For example, for receiver-$j$ and for the schedule $s$, this product message alphabet is the Cartesian product of $s_j$ copies of the original message alphabet $M_j$; hence the product message alphabet consists of

$M_j^{s_j}$ different tuples of length $s_j$. With this view point, we redefine the quantity $R_k$ (eq. (1) and (2) in Theorem 2.2), the coding rate for the receiver-$k$, as $R_k(s)$ thus emphasizing the dependence of *effective* message alphabet size on schedule $s$. Then

$$R_k(s) = \frac{\ln M_k^{s_k}}{N(s)} = s_k \frac{\ln M_k}{N(s)} = s_k R_k$$

Thus $R_k = R_k(s)$ for $s_k = 1$. Also, it is helpful to view schedule $s$ as a set of messages consisting of a total of $\sum_{j=1}^{J} s_j$ elements, of which the first $s_1$ messages are intended for receiver-1, the next $s_2$ messages are intended for transmission to receiver-2 and so on, and the last $s_J$ messages are intended for receiver-$J$. Let $\mathcal{P}(s)$ denote the set of all *non-empty* subsets of the schedule $s$, viewed as a set of messages. For future reference, we denote the random coding upper bound for the schedule $s$ and for the $j$th receiver by $\chi_j(s, N_j(s))$, where for a particular choice of $\rho$ and tolerable message decoding error probabilities $\{p_{ej}, 1 \leq j \leq J\}$, for $1 \leq j \leq J$ let $N_j(s)$ be the smallest positive integer such that $\chi_j(s, N_j(s)) \leq p_{ej}$. Then $p(\{\hat{m}_{j,j} \neq m_j\}) \leq p_{ej}$.

*Lemma 2.1:* Let $s' \in \mathcal{P}(s)$. Then, for $1 \leq j \leq J$, $N_j(s') \leq N_j(s)$. ∎

Since no closed form expression exists for $N_j(s)$, we derive an upper bound and a lower bound to $N_j(s)$ in Lemma 2.2. The notation that for $x > 0$ and $q > 0$, $\lceil x \rceil_q = \min(n \geq 1 : x \leq nq)q$ will be used in the following Lemma.

*Lemma 2.2:* Let $N_j(s)$ be the smallest positive integer such that $\chi_j(s, N_j(s)) \leq p_{ej}$. Then $N_j(s)$ can be bounded as shown below.

$$N_j(s) \geq \max_{j \leq k \leq J} \frac{\lceil -\ln p_{ej} + \rho s_k \ln M_k \rceil_{E_{o,X_k,Y_j}}}{E_{o,X_k,Y_j}}$$

$$N_j(s) \leq \max_{j \leq k \leq J} \frac{\lceil -\ln \frac{p_{ej}}{J-j+1} + \rho s_k \ln M_k \rceil_{E_{o,X_k,Y_j}}}{E_{o,X_k,Y_j}}$$

∎

Define $N(s) = \max_j N_j(s)$. Then $N(s)$ is the smallest positive integer such that for each $j$, $1 \leq j \leq J$, $\chi_j(s, N(s)) \leq p_{ej}$. In the following Lemma 2.3 we evaluate coding rates $R_k(s)$ under two asymptotic regimes.

*Lemma 2.3:* **(R1)** For $1 \leq j \leq J$ and an integer $M \geq 2$, let $M_j = M$. Then

$$\overline{R}_i(s) = \lim_{M \to \infty} R_i(s) = \min_{1 \leq j \leq J} \min_{j \leq k \leq J} \frac{s_i}{s_k} \frac{E_{o,X_k,Y_j}}{\rho}.$$

**(R2)** For $1 \leq j \leq J$ and an integer $t \geq 1$, let $s_j = t$. Let the positive integer vector $M = (M_1, M_2, \ldots, M_J)$ denote message alphabet cardinalities. Then

$$\overline{R}_i(M) = \lim_{t \to \infty} R_i(s) = \min_{1 \leq j \leq J} \min_{j \leq k \leq J} \frac{\ln M_i}{\ln M_k} \frac{E_{o,X_k,Y_j}}{\rho}$$

### III. QUEUEING-THEORETIC MODEL

In this section we derive a queueing-theoretic model for a $J$ receiver degraded broadcast channel, when requests for message transmission are randomly generated. This queueing model consists of $J$ queues, one for each receiver, and a single server whose service statistics depend on the state of the queues through the chosen scheduling policy.

Let successive maximum-likelihood decoding be used at each receiver to decode the respective received word. Consider a fixed schedule $s$ and suppose that a set of tolerable message decoding error probabilities $\{p_{ej}; 1 \leq j \leq J\}$ is given. The definition of service requirement that we consider for a message intended for any receiver is the smallest positive integer $N(s)$ (length of the code word that the transmitter transmits) such that $\chi_j(s, N(s)) \leq p_{ej}$. For the schedule $s$, we say that queue-$j$ receives a service quantum equivalent to $s_j$ units/slot; the total service quantum then is $\sum_{j=1}^{J} s_j$ units/slot. After receiving the signal transmission over $N$ channel uses, each receiver will decode the message intended for it. A few remarks on the definitions of service requirement and service quantum are in order. The service requirement of a message depends on the schedule of which the message is a component message. In other words, a message by itself *cannot* characterize service requirement for itself unless it is the only message to constitute the schedule. The amount of service quantum available to a receiver depends on the schedule.

Requests for message transmissions for each receiver are assumed to arrive at slot boundaries in batches. Let the random variable $A_j$, with finite moments $\mathbb{E}A_j$ and $\mathbb{E}A_j^2$, represent the number of messages destined for receiver-$j$ that arrive in any slot, with the pmf $\Pr(A_j = k) = p_j(k)$, $k \geq 0$. We assume that $\{A_j\}$ are independent random variables. Let $\mathbb{E}A = (\mathbb{E}A_1, \mathbb{E}A_2, \ldots, \mathbb{E}A_J) \in \mathbb{R}_+^J$. Let $\lambda_j$ denote the arrival rate of messages for the receiver-$j$. For channel bandwidth $W$, since each slot is of duration $\frac{1}{W}$, we have $\lambda_j = W\mathbb{E}A_j$.

Having defined service requirement and service quantum of a message for a given schedule, we are now in a position to analyze this message communication scheme with superposition encoding and successive decoding when requests for message transmission arrive at random times. We construct a discrete-time countable state space Markov-chain model of this communication system and then analyze for the stability ($c$-regularity [9]) of the model. The stability analysis consists of characterizing the stability region $\mathcal{R}(\omega) \in \mathbb{R}_+^J$ of message arrival rate vectors $\mathbb{E}A$ for each policy $\omega$ in a class of stationary "state-independent" scheduling policies by obtaining appropriate drift conditions for suitably defined Lyapunov functions of the state of the Markov chain. In particular, we prove that the Markov chain is $c$-regular by applying Theorem 10.3 from [9], and then show finiteness of the stationary mean number of messages in the system.

### IV. A GENERAL OUTER BOUND TO THE STABILITY REGION

In this section, we derive an outerbound to the region of message arrival rate vectors $\mathbb{E}A$ for which the Markov-chain model is positive recurrent and has finite stationary mean for the number of messages, for the class of stationary scheduling policies. Later, in Section V, we propose a class of stationary scheduling policies, called "state-independent" scheduling policies and denoted by $\Omega^{\mathsf{K}}$, and then prove that for any message arrival processes $\{A_j\}$ with $\mathbb{E}A_j$ inside the outerbound, there exists a scheduling policy $\omega \in \Omega^{\mathsf{K}}$ such that the Markov-chain model is positive recurrent and has finite stationary mean for the number of messages.

Consider message arrival processes $\{A_j; 1 \le j \le J\}$ and a stationary scheduling policy $\omega$ that schedules at most $\mathsf{K}$ messages for a joint message transmission. Let $\pi_{\mathsf{K}}(s)$ be a probability measure on $\mathcal{S}_{\mathsf{K}}$. Define
$\Psi_j = \sum_{\{s \in \mathcal{S}_{\mathsf{K}}: s_j > 0\}} \pi_{\mathsf{K}}(s) \frac{s_j}{N(s)}$ and the set

$$\mathcal{R}_{out} = \bigcup_{\pi_{\mathsf{K}}(s)} \{\beta \in \mathbb{R}_+^J : \beta_j \le \Psi_j\} \quad (3)$$

*Theorem 4.1:* Let the Markov chain $\{X_n, n \ge 0\}$ be positive recurrent and yield finite stationary mean for the number of messages in the system for the message arrival processes $\{A_j\}$ and the stationary scheduling policy $\omega$. Then $\mathbb{E}A \in \mathcal{R}_{out}$. ∎

## V. STABILITY FOR STATE-INDEPENDENT SCHEDULING POLICIES

In this section we define the class of stationary state-independent scheduling policies $\Omega^{\mathsf{K}}$, and then assert positive recurrence and finiteness of the stationary mean for the number of messages of the Markov-chain model for this class of scheduling policies. Formally, a policy in this class is defined by (i) a probability measure $\{p(s); s \in \mathcal{S}_{\mathsf{K}}\}$, and (ii) the mapping $\{\omega : \mathcal{X} \times \mathcal{S}_{\mathsf{K}} \to \mathcal{S}_{\mathsf{K}}\}$. In this paper, specification of a state independent scheduling policy $\omega$ and the probability measure $\{p(s); s \in \mathcal{S}_{\mathsf{K}}\}$ are equivalent. To implement a scheduling policy $\omega$, we first classify the incoming messages based on the particular schedule $s$ to be assigned to them.

For each message arrival destined for receiver-$j$, a schedule $s \in \{s \in \mathcal{S}_{\mathsf{K}} : s_j > 0\}$ is chosen randomly with the fixed probability measure defined later in (5) and the message is classified by assigning the class-$(j, s)$ to it. With this classification a message of class-$(j, s)$ will be scheduled to transmit *only* when the schedule $s$ gets chosen for transmission. One consequence of class sub-classification is that messages of class-$(j, s)$ will be required to use code words of length $N(s)$ for transmission, i.e., service requirement gets fixed. We first fix a scheduling policy $\omega = p(s)$ and then, in each time slot, a schedule $s$ is chosen from the set $\mathcal{S}_{\mathsf{K}}$, *independent* of the state $\alpha$, with probability $p(s)$. We constrain the operation of the system by requiring that there can be at most one on-going transmission [1] for any given schedule. The equivalent queueing model for any state-independent scheduling policy $\omega \in \Omega^{\mathsf{K}}$ will then consists of a number of queues, one for each message class. To define the state of the system, we keep track of the following information about each message class: for message class-$(j, s)$, let $n_{js}(\alpha)$ denote the number of fresh messages [2], $x_{js}$ the number of messages that are part of the on-going transmission, and $t_{js}$ the number of time-slots of transmission remaining for the on-going transmission. Define $\alpha_{js} = (n_{js}(\alpha), x_{js}, t_{js})$, the state information corresponding to message class-$(j, s)$ and then

$$\alpha = (\alpha_{js}; 1 \le j \le J, s \in \mathcal{S}_{\mathsf{K}}), \quad (4)$$

the state of the system.

Now we discuss implementation of the scheduling policy $\omega$. Suppose that the system is in state $\alpha$. Then the schedule to be selected for implementation in state $\alpha$ is a random variable and takes values in $\mathcal{S}_{\mathsf{K}}$. When trying to implement a schedule $s$ the following possibilities can occur:

1) For all of the message classes associated with the schedule $s$, there are no fresh messages present in the system; nor is there an ongoing transmission of schedule $s$. Then, no messages are scheduled in that state, and the system moves to next state as determined by the message arrival processes.
2) *No* on-going transmission of schedule $s$ is present in the system, and for at least one message class associated with the schedule $s$ there is at least one fresh message available. Then, a new joint message of schedule $s$ is scheduled, formed out of the fresh messages available with as many fresh messages of pertinent classes as are possible but not exceeding the respective maximum numbers specified by the schedule $s$.
3) There is an on-going transmission of schedule $s$ present in the system. Then that transmission is scheduled in that slot.

$\mathcal{X}$ is the countable set of state vectors $\alpha$ defined in (4). Let $V(\alpha)$ be a Lyapunov function defined on $\mathcal{X}$ and let $\mathcal{R}(\omega)$ denote the set of message arrival rate vectors $\mathbb{E}A$ for which the Markov chain $\{X_n, n \ge 0\}$ for the scheduling policy $\omega$ is positive recurrent and yields finite stationary mean for the number of messages of each class. Then we prove the following Lemma and two Theorems.

*Lemma 5.1:* For $\alpha \in \mathcal{X}$ and for message class-$(j, s)$, define $c_{js}(\alpha) = N(s)n_{js}(\alpha) + s_j t_{js}$. Next, define $c(\alpha) = 1 + \sum_{js} c_{js}(\alpha)$ and

$$V(\alpha) = \sum_{js} \frac{c_{js}^2(\alpha)}{2(p(s)s_j - \mathbb{E}A_{js}N(s))}.$$

Then, for the scheduling policy $\omega$, the Markov chain is $c$-regular if, for each message class-$(j, s)$, $\mathbb{E}A_{js}N(s) < p(s)s_j$. ∎

*Theorem 5.1:* Let, for at least one message class-$(j, s)$, $\mathbb{E}A_{js} > \frac{p(s)s_j}{N(s)}$. Then the Markov-chain $\{X_n, n \ge 0\}$ is transient. ∎

To prove Theorem 5.1, we show that for the Lyapunov function $V(\alpha) = 1 - \theta^{N(s)n_{js}(\alpha) + x_{js}(\alpha)t_{js}}$, there exists a value for $\theta$, $0 < \theta < 1$, for which $V(\alpha)$ satisfies the conditions for the theorem for transience [10]. Define $\mu_j = (\mu_{js}, s \in \mathcal{S}_\mathsf{K} : s_j > 0)$ be a splitting probability vector defined by

$$\mu_{js} = \frac{\frac{p(s)s_j}{N(s)}}{\sum_{\{s' \in \mathcal{S}_\mathsf{K}: s'_j > 0\}} \frac{p(s')s'_j}{N(s')}} \quad (5)$$

---

[1] A joint message for which at least one time-slot of transmission is complete and transmission for at least one more time-slot remains to be completed.

[2] We say that a message request is *fresh* if that message has not yet been scheduled for the first time, i.e., first code symbol of the corresponding code word is yet to be transmitted.

Then, given that a message arrives at queue-$j$, $\mu_{js}$ is the probability that the message request is assigned schedule $s$.

The sufficient condition for $c$-regularity of the Markov-chain $\{X_n, n \geq 0\}$ stated in Lemma 5.1 and the sufficient condition for transience stated in Theorem 5.1 together give the exact characterization of the stability region, as stated in the following theorem.

*Theorem 5.2:* For the scheduling policy $\omega$, the Markov chain $\{X_n, n \geq 0\}$ is

(a) positive recurrent and yields finite stationary mean for the number of messages, if, for each queue-$j$,

$$\mathbb{E}A_j < \sum_{\{s \in \mathcal{S}_\mathsf{K}: s_j > 0\}} \frac{p(s)s_j}{N(s)}, \quad \text{and}$$

(b) transient if, for at least one message class-$(j, s)$,

$$\mathbb{E}A_{js} > \frac{p(s)s_j}{N(s)}$$

■

Define $\psi_j = \sum_{\{s \in \mathcal{S}_\mathsf{K}: s_j > 0\}} p(s) \frac{s_j}{N(s)}$ and the set

$$\mathcal{R}\left(\Omega^\mathsf{K}\right) = \bigcup_{p(s) \in \Omega^\mathsf{K}} \{\beta \in \mathbb{R}_+^J : \beta_j < \psi_j\} \quad (6)$$

*Corollary 5.1:* For any given message arrival rate vector $\mathbb{E}A \in \mathcal{R}\left(\Omega^\mathsf{K}\right)$ there exists a scheduling policy $p(s) \in \Omega^\mathsf{K}$ such that the Markov chain is positive recurrent and yields finite stationary mean for the number of messages of each class. ■

From (3) and (6), we note [3] that $\mathcal{R}\left(\Omega^\mathsf{K}\right) = \mathcal{R}_{out}^o$. This observation essentially states that, if a stationary scheduling policy is stable for the message arrival processes $\{A_j; 1 \leq j \leq J\}$, then there exists a state-independent scheduling policy which makes the Markov-chain stable for the same message arrival processes $\{A_j; 1 \leq j \leq J\}$.

## VI. INFORMATION-THEORETIC CAPACITY REGION INTERPRETATION TO THE STABILITY REGION

In this section we give information-theoretic capacity region interpretation to the stability region of nat arrival rate vectors $\mathbb{E}\tilde{A}$. A formal statement of this interpretation is made in Theorem 6.1. Let $\tilde{A}_j = A_j \ln M_j$ denote the nat arrival random variable corresponding to message class-$j$. Then, the message system, for the fixed schedule $s$, is stable for nat arrival rates satisfying the following inequality: for receiver-$j$,

$$\mathbb{E}\tilde{A}_j < \frac{s_j \ln M_j}{N(s)} \quad (7)$$

[3]Interior of the set $A$ is denoted by $A^o$.

Inequality (7) follows trivially from Theorem 5.2. We remind the reader that $R_k(s) = \frac{s_j \ln M_j}{N(s)}$ denotes the maximum possible coding rate for receiver-$j$ under the schedule $s$ and that $R_k(s)$ under two asymptotic regimes was determined in Lemma 2.3. Define $\overline{R}(s) = (\overline{R}_1(s), \overline{R}_2(s), \ldots, \overline{R}_J(s))$ and the hypercube $\overline{\mathcal{R}}(s) \in \mathbb{R}_+^J$ defined by the vector $\overline{R}(s)$. Similarly, we define the vector $\overline{R}(M)$ and the hypercube $\overline{\mathcal{R}}(M) \in \mathbb{R}_+^J$ defined by the vector $\overline{R}(M)$. For the given joint distribution $q_J(x_J)q_{J-1}(x_{J-1}|x_J)\cdots q_1(x_1|x_2)p(y_1 y_2 \cdots y_J|x_1)$, let $\mathcal{I} = (I(X_J; Y_J), I(X_{J-1}; Y_{J-1}|X_J), \ldots, I(X_1; Y_1|X_2))$ denote the vector of average mutual informations and the hypercube $\mathcal{C}(\mathcal{I}) \in \mathbb{R}_+^J$ defined by the vector $\mathcal{I}$. We assert in the following Theorem 6.1 that, for any rate vector $r \in \mathcal{C}^o(\mathcal{I})$, there exists a schedule $s$ under regime **R1** and an $M$ under regime **R2** such that the Markov chain $\{X_n, n \geq 0\}$ under the respective regimes **R1** and **R2** with $\mathbb{E}\tilde{A} = r$, is stable. That is, the achievable asymptotic stable region of nat arrival rate vectors and the interior of the capacity region $\mathcal{C}(\mathcal{I})$ are identical.

*Theorem 6.1 (Capacity Interpretation):*

$$\bigcup_{\mathsf{K} \geq 1} \bigcup_{\{s \in \mathcal{S}_\mathsf{K}\}} \overline{\mathcal{R}}(s) = \bigcup_{M \in \mathbb{Z}_+^J} \overline{\mathcal{R}}(M) = \mathcal{C}^o(\mathcal{I})$$

■

Proof of Theorem 6.1 uses the following Lemma 6.1.

*Lemma 6.1:* Consider a $J$-receiver degraded broadcast channel represented as the Markov chain $X_J \to X_{J-1} \to \cdots \to X_1 \to Y_1 \to Y_2 \to \cdots \to Y_J$. Then, for $1 \leq j \leq J$ and $j \leq k \leq J$, $E_{o,X_k,Y_j} \geq E_{o,X_k,Y_k}$. ■